\documentclass[aps,prb,twocolumn,showpacs,superscriptaddress]{revtex4-1}

\usepackage{graphicx} % Include figure files

\begin{document}

\title
{Long ranged interactions in carbon atomic chains}
\author{S. Cahangirov}
\affiliation{UNAM-Institute of Materials Science and
Nanotechnology, Bilkent University, Ankara 06800, Turkey}
\author{M. Topsakal}
\affiliation{UNAM-Institute of Materials Science and
Nanotechnology, Bilkent University, Ankara 06800, Turkey}
\author{S. Ciraci} \email{ciraci@fen.bilkent.edu.tr}
\affiliation{UNAM-Institute of Materials Science
and Nanotechnology, Bilkent University, Ankara 06800, Turkey}
\affiliation{Department of Physics, Bilkent University, Ankara
06800, Turkey}

\date{\today}

\begin{abstract}
Based on first-principles calculations we revealed fundamental properties of infinite and finite size monatomic chains of carbon atoms in equilibrium and under an applied strain. Distributions of bond lengths and magnetic moments at atomic sites exhibit interesting even-odd disparity depending on the number of carbon atoms in the chain and on the type of saturation of carbon atoms at both ends. It was shown that, the $\pi$-bands of carbon atomic chains behave as a one dimensional free electron system. A local perturbation created by a small displacement of the single carbon atom at the center of a long chain induces oscillations of atomic forces and charge density, which are carried to long distances over the chain. These long ranged oscillations are identified as Friedel oscillations showing $1/r$ decay rate in one dimensional systems.
\end{abstract}

 \maketitle

\section{Introduction}
Carbon atomic chains are one-dimensional (1D) allotropic form of carbon atom, which has also allotropic forms in different dimensionalities, such as 3D diamond and graphite, 2D graphene, quasi 1D nanotube and quasi 0D fullerens. $sp^D$-hybrid orbitals are indigenous to the dimensionality (D=1,2,3) of these allotropic forms. The tetrahedrally coordinated $sp^3$-bonding stabilizes the open diamond structure. The $sp^2$-bonding together with $\pi$-bonding maintain the planar stability of honeycomb structure of graphene and attributes several exceptional physical properties.\cite{novoselov-nature-materials} Covalent bonding of $sp^{D=1}$ hybrid orbitals along the chain axis together with $\pi$-bonding of perpendicular $p_x$ and $p_y$ orbitals are responsible for the linear stability of the chain. Earlier carbon atomic chains and their functionalized forms (CACs) have been investigated intensively, despite the lack of consensus on whether they can really be synthesized.\cite{byl1998,abd2002,tongay_prl} These studies have predicted a wide range of interesting properties, which can make CACs a potential material for future nanotechnology applications.\cite{tongay_cond,dag_prb}

\begin{figure}
\includegraphics[width=8.5cm]{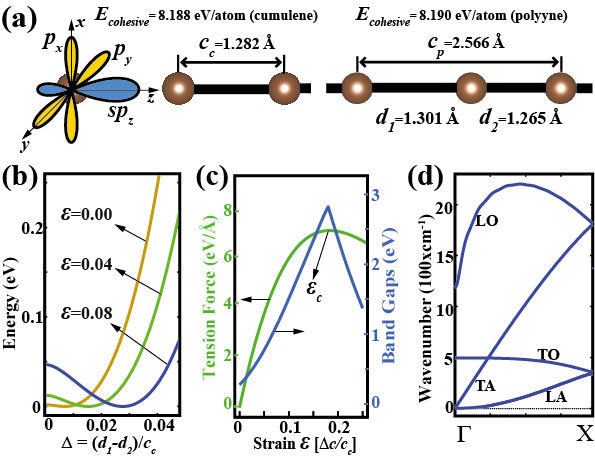}
\caption{(Color Online) (a) Schematic representations of orbitals and structure of cumulene (left) and polyyene (right). (b) The energy per unitcell of two carbon atoms versus the dimerization $\Delta = \delta/c_c$, that is defined as the ratio of displacement of one of the carbon atoms from the cumulene positions, $\delta=d_1-d_2$, to the lattice constant of cumulene, $c_c$. In each curve the minimum energy is set to zero. (c) Variations of band gap, $E_g$, and tension force, $F_T$, with applied strain, $\epsilon$. (d) Calculated phonon dispersions of polyyne structure.} \label{fig1}
\end{figure}

Linear CAC structures have either identical double bonds, called cumulene or alternating short "single" and strong "triple" bonds, called polyyne as described in Fig.~\ref{fig1}(a). While cumulene is metallic with a quantum ballistic conductance of $4e^2/h$ due to two degenerate, half-filled $p_{x}$ and $p_{y}$ bands crossing the Fermi level, it is vulnerable to Peierls instability.\cite{Peierls} Hence, through the displacement of alternating carbon atoms by $\delta~\cong$~0.018~\AA~the unitcell is doubled and a band gap of 0.32 eV at the edge of Brillouin zone (BZ) is opened to lower the total energy per atom by 2 meV. While segments of polyynes terminated with H atoms at both  ends (H-C$_{n}$-H), have been produced\cite{hcnh,polycan} up to considerable lengths ($n$=20), cumulene production is relatively difficult due to their frailty. Small cumulene chains terminated by H$_{2}$, (H$_{2}$-C$_{n}$-H$_{2}$) groups have been synthesized.\cite{h2cnh2,cumucan} Freestanding CACs from graphene flakes were produced\cite{iijima} by using energetic electron irradiation inside a transmission electron microscope (TEM). Concomitantly, it was demonstrated that not only CACs, but also SiACs and BNACs can be derived from their corresponding honeycomb structure under uniaxial tensile stress in the plastic deformation range.\cite{topsakal} Much recently, polyyne structure consisting of 44 carbon atoms have been produced.\cite{polchem}

In this work, we predict that the even-odd disparities are attained in the distributions of bond lengths and atomic magnetic moments of finite size CACs depending on the type of saturation of carbon atoms at both ends. Even more remarkable is that a local perturbation through the displacement of a single chain atom creates atomic force and charge density oscillations, which propagate to long distances in the chain. We show that, these long ranged couplings in CACs can be explained in terms of Friedel theory.\cite{friedel}

\section{Methods}
Our predictions are obtained from the state-of-the-art first-principles plane wave calculations carried out within the density functional theory (DFT) using PAW potentials.\cite{paw} The exchange correlation potential is approximated by generalized gradient approximation (GGA) using PBE functional. A plane-wave basis set with kinetic energy cutoff of 450 eV is used. All CACs are treated by supercell geometry and atomic positions and lattice constants are optimized by using the conjugate gradient method, where the total energy and atomic forces are minimized. \cite{vasp} The vacuum separation between the  CACs in the adjacent unit cells is taken to be at least 10~\AA. The convergence for energy is chosen as 10$^{-5}$ eV between two  steps, and the maximum Hellmann-Feynman forces acting on each atom  is less than 0.01 eV/\AA{} upon the ionic relaxation. Phonon dispersions  were obtainned using the force constant method with forces calculated in a ($40 \times 1 \times 1$) supercell.\cite{alfe}

\begin{figure*}
\includegraphics[width=15cm]{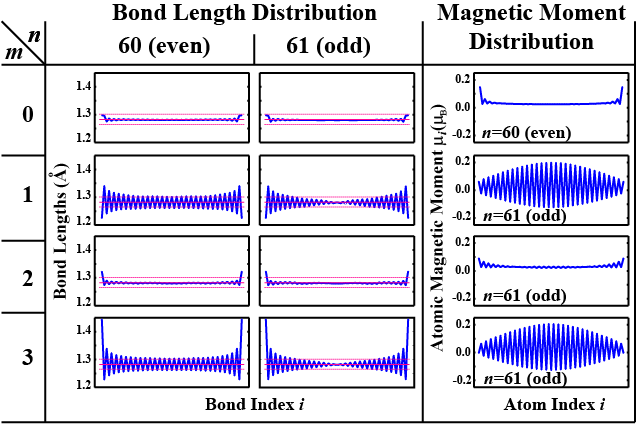}
\caption{(Color Online) The even-odd disparity: Distribution of C-C bond lengths and magnetic moments, $\mu_i$, at chain atoms of a finite CAC including $n$ C atoms, which is saturated by $m$ hydrogen at both ends, namely (H$_m$-C$_n$-H$_m$). $m$=0 corresponds to the bare CAC with free ends. H-saturated ($m$=1,2,3) CACs with even $n$ and bare CACs with odd $n$ have nonmagnetic ground state. All other CACs have total magnetic moment of 2$\mu_B$ distributed on the atomic cites as shown in the right panel. The dotted lines are a guide to the equilibrium bond distances of cumulene and polyyne structures.} \label{fig2}
\end{figure*}

\section{Infinite Carbon Atomic Chains}
The variation of energy as a function of dimerization $\Delta$ in Fig.~\ref{fig1}(b), shows that cumulene structure corresponds to a metastable state which transforms to polyyne structure without facing any energy barrier. This instability is also reflected in the phonon modes of cumulene structure. The longitudinal acoustic modes of cumulene attain imaginary frequencies near the zone boundary. The analogy of this transition can be found in graphene, where 1.5 fold bonds of the equilateral hexagons break into single and double bonds. It was shown that, if the non-local part of the exchange-correlation is taken into account through hybrid functionals, the high symmerry phase of graphene becomes considerably less stable.\cite{laz,gru,maf} Thus, we have performed a structural optimization of cumulene and polyyne structures also using HSE, B3LYP and PBE0 hybrid functionals, as they are implemented in VASP.\cite{hyb1,hyb2,hyb3} Indeed, the energy difference between the cumulene and polyyne structures, which is 2 meV according to PBE functional, is increased to 47 meV, 80 meV and 104 meV when HSE, B3LYP and PBE0 hybrid functionals are used, respectively. Interestingly, while the ground state of infinite CAC is polyyne and hence cumulene transforms to polyyne, cumulene structure, by itself, can be stabilized by charging. We deduced interesting effects of charging on infinite and finite CACs, which is beyond the scope of the present study.

In Fig.~\ref{fig1}(b) one can see that applying stress to the CACs enhances dimerization both energetically and spatially. As the lattice constant is increased by 8\% the energy difference between polyyne and cumulene structures is increased to 40 meV as compared to 2 meV difference in the absence of stress. Also the spatial dimerization relative to the lattice constant is 3 times higher in 8\% strained structure than that in the unstrained one. The increase of dimerization with strain results in the increase of the band gap, as shown in Fig.~\ref{fig1}(c). The band gap reaches its maximum value of 2.87 eV (which is an order of magnitude higher than that in the unstrained case) at a critical strain $\epsilon_c$=0.18. At this critical strain, the band gap switches from direct to indirect, since the minimum of the second conduction band starts to dip below the first conduction band. This rapid change of band gap with strain is interesting and can make CACs a potential candidate for strain gauge type nanodevice applications.

Fig.~\ref{fig1}(c) also presents the variation in the tension force $F_T$ with strain. The tension force is defined as $F_T=-\partial E_T/\partial c_p$, where $E_T$ is the total energy per unit cell. $F_T$ curve starts with a linear region at small strain values, which corresponds to the elastic regime. Following this region, the slope of this curve, which is proportional to the sound velocity, starts to decrease.\cite{cohen} This lasts until the sound velocity drops to zero where the tension curve reaches its maximum. This is the critical point occurring at $\epsilon_c$, beyond which CAC cannot sustain the long wavelength perturbations.

The phonon dispersions of unstrained polyyne is presented in Fig.~\ref{fig1}(d). Degenerate transversal phonon branch of polyyne structure mimics the folded version of that of cumulene, except a small gap which separates this branch into acoustical and optical modes. High frequency (wave number) longitudinal branch of polyyne also remind the folded version of that of cumulene but now there is a dramatic difference in a sense that imaginary frequencies at long wavelength disappear. The transversal acoustic mode loses its quadratic behavior near the $\Gamma$ point under strain. Longitudinal acoustic and optical modes, however, change dramatically with strain. The dispersion of the optical branch decreases with increasing strain. In the limit of very high strains where individual carbon dimers do not interact, this mode is expected to converge to a flat line with frequency corresponding to the vibrational motion of an isolated carbon dimer. The sound velocity related with the slope of the longitudinal acoustic mode decreases with increasing strain. The imaginary frequencies appear near the $\Gamma$-point beyond $\epsilon_c$, which indicates the onset of instability under the long wavelength perturbations.

The phonon modes are derived by using the direct method where the force constants are calculated in a ($40 \times 1 \times 1$) unitcell comprising 80 atoms.\cite{alfe} Such a long unitcell comprising 80 atoms was necessary because force constants in longitudinal directions decay rather slowly. On the other hand, the transversal force constants decay rapidly, which leads to appearance of quadratic terms in the phonon dispersions of transversal modes near the center of BZ. It is interesting to compare the decay of force constants of CACs with those of graphene. In the honeycomb structure the magnitude of the out of plane (transversal) and in-plane force constants of 5$^{th}$ neighbor, which is 4.3~\AA~apart, is about $1/100$ and $1/50$ of the nearest neighbor force constants, respectively.\cite{dubay,graphene} Similar to graphene, in CACs, the transversal force constant is reduced to $\sim 1/100$ at the 4$^{th}$ neighbor which is 5.1~\AA~apart. However, the longitudinal force constant of CACs is reduced to about $1/50$ only at the 22$^{nd}$ neighbor, which is 28.2~\AA~apart. This slow decay in the magnitudes of longitudinal force constants of CACs is accompanied with sign oscillations starting from 3$^{rd}$ neighbor. The slow decay of longitudinal force constants in CACs indicate long ranged nature of specific interactions, which is the principal subject of our study and will be treated in the forthcoming parts.

\begin{figure*}
\includegraphics[width=15cm]{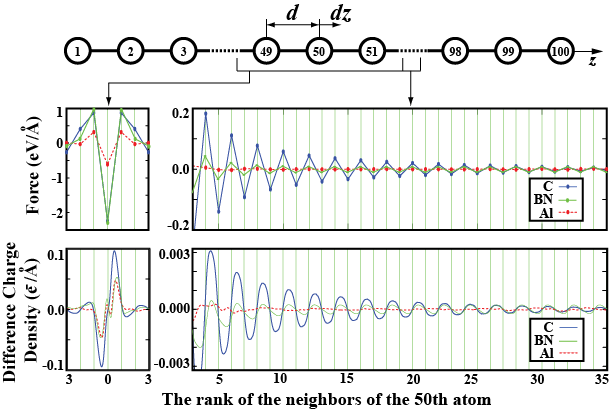}
\caption{(Color Online) Forces and difference charge density perturbations formed in finite C$_{100}$, B$_{50}$N$_{50}$ and Al$_{100}$ atomic chains due to a small longitudinal displacement, $\delta z$, of the 50$^{th}$ atom near the center of the chain. The upper panel is a schematic representation of the geometry of the system. The middle panel presents force distribution for three nearest neighbors (left) and extensions up to 35$^{th}$ neighbor. In the similar manner the bottom panel presents the difference charge density perturbations. Atomic magnetic moments (which are not shown in the figure) also behave similarly, and have long ranged oscillations in CAC.} \label{fig3}
\end{figure*}

\section{Long Ranged Interactions in Finite Carbon Atomic Chains}
Infinite CACs cannot exist in reality, but the properties of long finite CACs are expected to converge to those of infinite ones. There are two new parameters which can affect the properties of finite CACs dramatically. These are the number of carbon atoms $n$ forming the chain and the \textit{end effects}. To simulate the end effects we either leave both ends of CACs bare or passivate each of them by one, two or three hydrogen atoms. A similar analysis, which confirms our results, was performed earlier for only bare CACs and those passivated with one hydrogen from both edges.\cite{acsnano} We also expect that the CAC passivated with two (three) hydrogen atoms from both ends to behave similarly to CACs attached between graphene sheets (diamond blocks).\cite{milani} Figure~\ref{fig2} presents the distribution of bond lengths and atomic magnetic moments of finite CACs having $n$=60 and 61 carbon atoms saturated by $m$ hydrogen atoms from both ends. It turns out that for a given $m$ the structural and magnetic properties can be classified in two classes depending on whether $n$ is even or odd. This means that the distribution of bond length and magnetic moments of (H$_2$-C$_{60}$-H$_2$) and (H-C$_{61}$-H) structures, for example, is similar to that of (H$_2$-C$_{100}$-H$_2$) and (H-C$_{101}$-H), respectively. These trends leading to even-odd disparity are confirmed for chains with $n=20, 21, 60, 61, 100, 101$ and for $m=0, 1, 2, 3$.

When edges are passivated by only one H atom, the C atoms at both edges make single bond with these H atoms. This forces the type of C-C bond to the adjacent chain atom to be triple. Then the next C-C bond is forced to be single and so on. This alternation of triple and single bonding is also reflected to the short and long bond length alternations. A situation contrary to this occurs, when the edges are passivated with three H atoms. This time H atoms are arranged tetrahedrally and the outermost C-C bonds at both edges is forced to be single so that the bond length alternation starts with a longer bond. These structures are so similar that the C-C bonds in the structure passivated with one H atom remain unchanged if these H atoms passivating both ends are replaced by CH$_3$ groups. The bond length alternations in these structures are enhanced at the edges. For even values of $n$, the middle parts acquire a bond length alternation between two values corresponding to the bond lengths of polyyne. For odd values of $n$, however, the bond length alternation originated from both edges compete at the middle parts and the pattern presented in Fig.~\ref{fig2} is formed.

When the ends are passivated by two H atoms, the type of C-C bonds at the ends becomes double. This time all other C-C bonds acquire the same type. As a result, for sufficiently long CACs, the bond length alternation in the middle parts become negligible while the bond lengths are very close to that found in cumulene. Carbon atomic chains making two bonds with cone-terminated carbon nanotubes from both edges were shown to have cumulene type bonding, which corroborates our findings.\cite{nanotube} Following the similarity observed between the CACs passivated by single and triple H atoms, the bare CACs and CACs passivated by two H atoms have similar atomic structure. That is, bare CACs also have double bonds between C atoms and have negligible bond length alternations in the middle parts. It is remarkable that an atom at the center of a chain as long as 100 C atoms, is affected by the type of passivation at its edges.

Similar trends are obtained in the distribution of atomic magnetic moments, where their values oscillate until long distances. Since the effect of a specific type of saturation occurring at both ends of a long chain can be carried over to distant neighbors, resulting even-odd disparities imply a long ranged nature of couplings in CACs.

We now consider the aforementioned atomic force and difference charge density oscillations of bare C$_{100}$, B$_{50}$N$_{50}$, B$_{100}$ and Al$_{100}$ chains generated by a slight displacement of the 50th atom. Despite its zigzag metallic ground state, \cite{sen} here we consider linear Al-chain for the sake of comparison. After the structural relaxation, all atomic positions are kept fixed except the 50$^{th}$ atom, which was displaced in longitudinal direction by $\delta z$ taken to be 0.02~\AA{} for C$_{100}$, B$_{50}$N$_{50}$, B$_{100}$ and 0.04~\AA{} for Al$_{100}$ chain. Difference charge density is obtained by subtracting the charge density of the perturbed structure from that of the unperturbed one, which is averaged in planes perpendicular to the chain axis. In the left panel of Fig.~\ref{fig3}, the force and linear charge density perturbations of only three nearest neighbors of the 50$^{th}$ atom is shown. The right panel of the Fig.~\ref{fig3} presents the extensions of these perturbations up to the 35$^{th}$ neighbor of the 50$^{th}$ atom. One can see that these extensions are negligible and show no obvious pattern in Al chains. Similar pattern is also seen in Boron atomic chains. In CACs, however, the perturbation extensions are long ranged and exhibit a decaying oscillatory behavior. The envelope of these oscillatory extensions fits to a $1/r$ decay rate ($r$ being the distance to the 50th atom) for both force and linear charge density perturbations. Similar pattern is also seen in B$_{50}$N$_{50}$ chain, but both force and charge density oscillations are weaker compared to that of C$_{100}$.

\section{Discussions and Conclusions}

The above oscillatory decay in the linear charge density perturbation provokes an association with so called Friedel oscillations.\cite{friedel} Inserting an impurity charge to an electron gas results in the accumulation of electronic charge, which screens the Coulomb potential introduced by the impurity. Since the wave  vector of the electron gas is limited by its density, the accumulated charge acquires decaying Friedel oscillations. For sufficiently large separations, $r$, from the impurity, the Friedel oscillations are proportional to the function $sin(2 k_{F} r)/r^{D}$, where $k_{F}$ and $D$ stand for the Fermi wave vector and dimensionality of the electron gas system, respectively.\cite{giu1} Recently it was shown that, upon inclusion of an impurity in a two-dimensional graphene structure, where $D=2$, the charge density oscillations proportional to $sin(2 k_{F} r)/r^{2}$ are obtained.\cite{arxiv} In case of CAC, one can think of the displacement of the 50$^{th}$ atom near the center of the chain as the insertion of two impurity charges having opposite signs.\cite{grosu} Since the carbon chain system is one dimensional, the Friedel theory can explain the $1/r$ decay rate.\cite{giu2} In cumulene structure the double degenerate $\pi$-bands are half filled, so the Fermi wave vector is $k_{F} = \pi /2c_{c}$. In polyyne structure bands are folded and the Brillouin zone is halved and the Fermi wave vector of filled $\pi$-bands is $k_{F} = \pi /c_{p}$. Thus the periodicity of observed charge density oscillations fits the Friedel theory for both polyyne and cumulene structures (and also for BN chains).

\begin{figure}
\includegraphics[width=8.5cm]{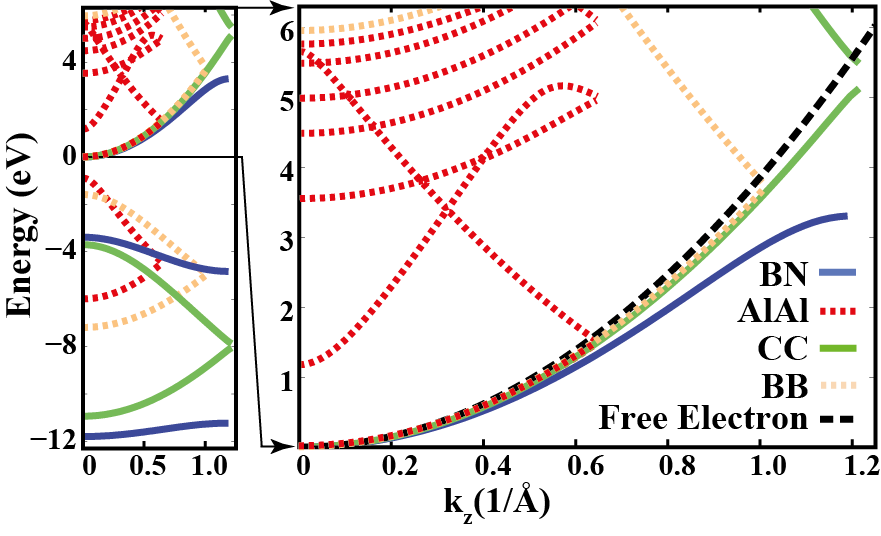}
\caption{(Color online) Energy band structures of AlAl, BB, BN and polyyne chain structures. A closer view of $\pi$-bands together with the dispersion profile of one-dimensional free electron gas system is presented in the right panel. Zero of energy is set to the minima of $\pi$-bands for each structure. For the sake of clarity, the bands of the chain structures beyond the first Brillouin zone in the extended scheme are not shown.} \label{fig4}
\end{figure}

One should note that, Friedel theory holds for free electron like systems. In our case, the $\pi$-bands of atomic chains are expected to behave as free electron systems, since their charge density have nodes on the axis where the ions lie. Figure~\ref{fig4} presents the band structure of AlAl, BB, BN and polyyne chain structures. All structures are considered in a unitcell having two atoms for the sake of comparison. The left panel presents the whole band profile, while the right panel zooms to the $\pi$-bands of chain structures and compare them with free electron dispersion. As seen in the right panel of Fig.~\ref{fig4}, the $\pi$-states of all chain structures considered here have a dispersion profile which is very similar to that of a free electron. In fact the effective masses of $\pi$-bands of BB, CC, AlAl and BN chain structures exceed the mass of free electron by only $2 \%$, $7 \%$, $8 \%$ and $17 \%$, respectively. Here the effective masses are calculated from the curvature of the $E(k)$ profile around the $\Gamma$ point.

BB and AlAl chains both have six valence electrons in a unitcell consisting of two atoms. The folded double degenerate $\pi$-bands of these structures are half filled, which result in $k_{F} = \pi /4c$. Considering that, the $\pi$-band dispersion of BB chain is closer to free electron dispersion compared to that of polyyne, one expects the BB and AlAl chains to have Friedel oscillations with a periodicity of $4c$. However, there is no such profile present in force or charge density oscillations of AlAl chains, as seen in Fig.~\ref{fig3}. We think that, in AlAl and BB chains the Friedel oscillations, which is a feature of a free electron systems, is suppressed by ionic potential. In CC and BN chains the oscillations have a periodicity of $2c$, which resultes in very small charge density perturbations near the atomic sites. In AlAl and BB chains, however, the charge density oscillations with a period of $4c$ are expected to have maxima on atomic sites. This enhances the contribution of ionic potential, so that, $\pi$-bands deviate from free electron like behavior. As a result, the charge density perturbations, being a feature of free electron like system, are suppressed and the system finds a self consistent energy minimum with no Friedel oscillations. Note that, force and charge density oscillations are weaker in BN chains compared to that of CACs. This is because, the $\pi$-bands of BN chains considerably deviate from free electron behavior, as seen in Fig.~\ref{fig4}.

The force distribution observed in Fig.~\ref{fig3} can be related to the charge density oscillations; as mentioned before, in the absence of any perturbation, the minimum of energy is met by symmetric charge densities around the atomic sites. As the charges start to move in one direction, the atomic sites start to feel a force in the same direction. To the first order, these two events are directly proportional and this explains the oscillating $1/r$ decay of the atomic forces. Note that, the 50$^{th}$ atom and its neighbors feel the elastic forces generated by displacement. Thus, they are not in the same direction as the charge accumulation.

In conclusion, a local displacement of an atom in carbon atomic chains creates long ranged oscillations of atomic forces and charge density. These remarkable and fundamental features of carbon chains are explained in terms of Friedel Theory developed for one-dimensional free electron system.

We thank A. Virosztek for helpful discussion. This work is partially supported by TUBA. Part of the computations have been provided by UYBHM at Istanbul Technical University through a Grant No. 2-024-2007.


\begin{thebibliography}{15}

\bibitem{novoselov-nature-materials}
A. K. Geim, and K. S. Novoselov, Nature Materials \textbf{6}, 183 (2007).

\bibitem{byl1998}
E. J. Bylaska, J. H. Weare, and R. Kawai, Phys. Rev. B \textbf{58}, R7488 (1998).

\bibitem{abd2002}
A. Abdurahman, A. Shukla, M. Dolg, Phys. Rev. B \textbf{65}, 115106 (2002).

\bibitem{tongay_prl}
S. Tongay, R. T. Senger, S. Dag, and S. Ciraci, Phys. Rev. Lett. \textbf{93}, 136404 (2004).

\bibitem{tongay_cond}
S. Tongay, S. Dag, E. Durgun, R. T. Senger, and S. Ciraci, J. Phys. Cond. Matt.  \textbf{17}, 3823 (2005).

\bibitem{dag_prb}
S. Dag, S. Tongay, T. Yildirim, R. T. Senger, C. Y. Fong, and S. Ciraci, Phys. Rev. B \textbf{72}, 155444 (2005).

\bibitem{Peierls}
R. E. Peierls, \textit{Quantum Theory of Solids} (Oxford University Press, New York, 1955), p. 108.

\bibitem{hcnh}
T. Pino, H. Ding, F. Guthe, and J. P. Maier, J. Chem. Phys. \textbf{114}, 2208 (2001).

\bibitem{polycan}
S. Eisler, A. D. Slepkov, E. Elliott, T. Luu, R. McDonald, F. A. Hegmann, and R. R. Tykwinski, J. Am. Chem. Soc. \textbf{127}, 2666 (2005).

\bibitem{h2cnh2}
X. Gu, R. I. Kaiser, and A. M. Mebel, Chem. Phys. Chem. \textbf{9}, 350 (2008).

\bibitem{cumucan}
S. Hino, Y. Okada, K. Iwasaki, M. Kijima, and H. Shirakawa, Chem. Phys. Lett. \textbf{372}, 59 (2003).

\bibitem{iijima}
C. H. Jin, H. P. Lan, L. M. Peng, K. Suenaga, and S. Iijima, Phys. Rev. Lett. \textbf{102}, 205501 (2009).

\bibitem{topsakal}
M. Topsakal and S. Ciraci, Phys. Rev. B \textbf{81}, 024107 (2010).

\bibitem{polchem}
W. A. Chalifoux and R. R. Tykwinski, Nature Chem. \textbf{2}, 967 (2010).

\bibitem{friedel}
J. Friedel, Phil. Mag. \textbf{43}, 153 (1952).

\bibitem{paw}
P. E. Bl\"ochl, Phys. Rev. B \textbf{50}, 17953 (1994).

\bibitem{vasp}
G. Kresse, J. Furthmuller, Phys. Rev. B \textbf{54}, 11169 (1996).

\bibitem{alfe}
D. Alf\`e, Comput. Phys. Commun. \textbf{180}, 2622 (2009).

\bibitem{laz}
M. Lazzeri, C. Attaccalite, L. Wirtz and Francesco Mauri, Phys. Rev. B, \textbf{78}, 081406(R) (2008).

\bibitem{gru}
A. Grüneis, J. Serrano, A. Bosak, M. Lazzeri, S. L. Molodtsov, L. Wirtz, C. Attaccalite, M. Krisch, A. Rubio, F. Mauri, and T. Pichler, Phys. Rev. B, \textbf{80}, 085423 (2009).

\bibitem{maf}
D. L. Mafra, L. M. Malard, S. K. Doorn, Han Htoon, J. Nilsson, A. H. Castro Neto, and M. A. Pimenta, Phys. Rev. B, \textbf{80}, 241414(R) (2009).

\bibitem{hyb1}
J. Heyd, G. E. Scuseria and M. Ernzerhof, J. Chem. Phys. \textbf{118}, 8207 (2003).

\bibitem{hyb2}
J. Paier, M. Marsman, K. Hummer, G. Kresse, I. C. Gerber and J. G. \'Angy\'an, J. Chem. Phys. \textbf{124}, 154709 (2006); \textbf{125}, 249901(E) (2006).

\bibitem{hyb3}
J. Paier, M. Marsman, and G. Kresse, J. Chem. Phys. \textbf{127}, 024103 (2007).

\bibitem{cohen}
F. J. Ribeiro and M. L. Cohen, Phys. Rev. B, \textbf{68}, 035423 (2003).

\bibitem{dubay}
O. Dubay and G. Kresse, Phys. Rev. B \textbf{67}, 035401 (2003).

\bibitem{graphene}
It was also deduced that, the interaction between two adatoms adsorbed on graphene (like C, Si and Ge pairs) is long ranged. See for example; E. Akt\"urk, C. Ataca and S. Ciraci, Appl. Phys. Lett. \textbf{96}, 123112 (2010); C. Ataca, E. Akt\"urk, H. \c Sahin and S. Ciraci, J. Appl. Phys. \textbf{xx}, (2010).

\bibitem{acsnano}
X. F. Fan, L. Liu, J. Y. Lin, Z. X. Shen and J. L. Kuo, ACS Nano \textbf{3}, 3788 (2009).

\bibitem{milani}
L. Ravagnan, N. Manini, E. Cinquanta, G. Onida, D. Sangalli, C. Motta, M. Devetta, A. Bordoni, P. Piseri and P. Milani, Phys. Rev. Lett. \textbf{102}, 245502 (2009).

\bibitem{nanotube}
H. E. Troiani, M. Miki-Yoshida, G. A. Camacho-Bragado, M. A. L. Marques, A. Rubio, J. A. Ascencio, and M. Jose-Yacaman, Nano Lett. \textbf{3}, 751 (2003).

\bibitem{sen}
P. Sen, S. Ciraci, A. Buldum, I. P. Batra, Phys. Rev. B \textbf{64}, 195420 (2001).

\bibitem{giu1}
G. F. Giuliani and G. Vignale, \textit{Quantum Theory of the Electron Liquid} (Cambridge University Press, Cambridge, England, 2005).

\bibitem{arxiv}
\'A. B\'acsi and A. Virosztek, arXiv:1009.2905.

\bibitem{grosu}
I. Grosu and L. Tugulan, J. Supercond. Nov. Magn. \textbf{21}, 65 (2008).

\bibitem{giu2}
G.F. Giuliani, G.Vignale, T. Datta, Phys. Rev. B \textbf{72}, 033411 (2005).

\end{thebibliography}
\end{document}